\documentclass[preprint,12pt,authoryear]{elsarticle}

\usepackage{amssymb}
\usepackage{url}
\usepackage{color}
\usepackage{comment} 
\usepackage{booktabs}
\usepackage{multirow}
\usepackage[normalem]{ulem}

\newcommand{\smalltilde}{\scriptsize$\sim$\normalsize}

\newenvironment{itemizesmall}
{\footnotesize\begin{list}{$\bullet$}{
\leftmargin=5mm
\labelwidth=3mm
\labelsep=3mm
}}{\end{list}\normalsize}

\journal{International Journal of Human Computer Studies}

\begin{document}

\begin{frontmatter}

\title{Submitting surveys via a conversational interface: an evaluation of user acceptance and approach effectiveness}

\author{Irene Celino and Gloria {Re Calegari}}

\address{Cefriel -- Politecnico di Milano, viale Sarca 226, 20126 Milano, Italy}

\begin{abstract}
Conversational interfaces are currently on the rise: more and more applications rely on a chat-like interaction pattern to increase their acceptability and to improve user experience. Also in the area of questionnaire design and administration, interaction design is increasingly looked at as an important ingredient of a digital solution. For those reasons, we designed and developed a conversational survey tool to administer questionnaires with a colloquial form through a chat-like Web interface.

In this paper, we present the evaluation results of our approach, taking into account both the user point of view -- by assessing user acceptance and preferences in terms of survey compilation experience -- and the survey design perspective -- by investigating the effectiveness of a conversational survey in comparison to a traditional questionnaire. We show that users clearly appreciate the conversational form and prefer it over a traditional approach and that, from a data collection point of view, the conversational method shows the same reliability and a higher response quality with respect to a traditional questionnaire. 
\end{abstract}


\end{frontmatter}

\section{Introduction}\label{sec:intro}
In the last few years, with the renewed interest on artificial intelligence and machine learning, autonomous agents and chatbots are experiencing a new popularity. The availability of intelligent services at our fingertips -- being it voice search on mobile or personal assistants at home (like Siri, Google Assistant, Alexa, etc.) -- enormously increased the interest around the so-called conversational interfaces. The goal is to provide the user or customer with a natural interaction pattern that resembles the human dialogue, even when the counterpart is a computer. This is also in line with the continuously increasing use of messaging applications, especially on mobile devices. 

On the other hand, in the field of questionnaire design and research, a growing attention is devoted to how a survey is administered to its users for data collection. By having a look at the major online survey tools and platforms, we notice that user context and user experience have gained traction and importance in the market offering. Just to make a couple of examples, SurveyMonkey\footnote{Cf. \url{https://www.surveymonkey.com/} (last visited: 2020/01/20).} gives the possibility to share a survey through a Facebook Messenger channel and Typeform\footnote{Cf. \url{https://www.typeform.com/} (last visited: 2020/01/20).} focuses all its competitive differentiation on user interface and interaction.

Therefore, the introduction of conversational approaches in questionnaire design can be considered an interesting trend, based on the intuition that a more natural interaction with the survey tool can be an effective incentive for users to participate to data collection. Still, to the best of our knowledge, this is one of the first studies performed to get an experimental proof of such a hypothesis and the first one that does not require any intelligent agent technology. 

The purpose of this paper is thus to provide evidence of the actual advantages that can be obtained by adopting a conversational approach in survey design and administration. Our goal is to show that, on the one hand, a conversational survey is well perceived or even preferred by users because of its more engaging user experience and, on the other hand, that a conversational way of administering a questionnaire is a reliable survey methodology potentially leading to a higher response quality and, as such, can be used as an effective and possibly superior substitute to a more traditional method, without the need to train a chatbot. 

The remainder of the paper is organized as follows: Section~\ref{sec:related} presents related work and Section~\ref{sec:tool} gives an overview of our conversational survey toolkit; the experimental setup of our evaluation is illustrated in Section~\ref{sec:exp}, while the collected results are explained in details in Section~\ref{sec:ux} in relation to user experience and in Section~\ref{sec:meth} for what regards the method effectiveness; finally, in Section~\ref{sec:concl} we draw some conclusions.

\section{Related work and motivation}\label{sec:related}
Our research on conversational surveys moves from different areas: on the one hand, we build on the large and vast body of knowledge about survey methods, both within the quantitative and qualitative research areas, also stemming from the introduction of computer-mediated communication; we also leverage all the literature on chatbots, intelligent agents and virtual agents, which is experiencing a recent increased attention, several years after its beginnings in the '60s; finally, we stress the importance of user experience in the actual engagement of survey compilers.

\subsection{Quantitative and qualitative survey research methods}
The differences, commonalities and interplay between quantitative and qualitative research methods have been explored for several decades. 

Qualitative research is an observation method based on non-numerical information, usually carried out by means of interviews. Quantitative research methods, on the other hand, investigate phenomena via statistical, mathematical, or computational techniques applied to  numerical data.

\cite{schober1997does} propose a discussion about standardized vs. flexible interviewing (i.e., reading questions verbatim vs. providing support to understanding, named ``conversational interviewing'') and analyse their impact on answer accuracy and on interview duration.

\cite{gobo2011back} investigates the differences between open and closed questions, the challenges to make an interview closer to a conversation, while avoiding all the biases for closed answers; he also suggests the adoption of the technique proposed by \cite{galtung1967theory}, named ``conversational survey'', as a mixed qualitative-quantitative method to leave the interviewer free in her method, while balancing the pros and cons of the various techniques.

In general, when approaching interviews, proper care should be given to the cognitive processes that both interviewers and respondents experience. \cite{ongena2007model} provide an explanation of such cognitive processes and illustrates the challenges, spanning from question formulation, question interpretation, answer retrieval, response formatting/coding and finalization. \cite{tourangeau2018survey} proposes the CASM model (Cognitive Aspects of Survey Methodology) to take care of the cognitive process involved in survey response.

With respect to quantitative research, an analysis of the challenges in questionnaire design is provided by \cite{krosnick2018improving}: the author illustrates the cognitive steps when answering a question (and the risk of satisficing), the violation of the basic conversational rules (like having multiple questions to measure the same construct), the characteristics of open and closed questions and their impact on validity and reliability (including risks of salience, framing, obscurity, rating scale range, scale labeling), as well as the issues related to question wording and language.

Our research effort is aimed to provide a set of tools to set up a quantitative questionnaire, but involving characteristics and advantages of qualitative research methods. Our concept of ``conversational survey'' is a questionnaire disguised as a conversation, which proposes qualitative answer options which however are automatically quantitatively coded for a numerical analysis.

\subsection{Face-to-face, paper-based and computer-mediated survey}

With the advent of digital technologies, computer-based approaches were introduced and compared to classical methods which were based on personal interaction and paper-based interviewing. Acronyms like CAPI, CASI, CATI and CAWI (computer-assisted personal/self/telephone/Web interviewing) started to become popular since the 80s and to be adopted for longitudinal studies, like the case in \cite{thornberry1991use}. 

The advantages with respect to face-to-face interviewing and paper-and-pen methods were demonstrated in particular with respect to an improved quality of the collected data, as investigated by \cite{sebestik1988initial} and \cite{beckenbach1995computer}; another interesting effect was identified in relation to an increase in self-disclosure of personal information, especially in conjunction with anonymity, as shown by \cite{weisband1996self}, \cite{van2000comparison} and \cite{joinson2001self}; in some specific cases, like suicidal patients, a computer-mediated approach proved to be even more advisable that an interview with a physician, leading also to a better prediction of suicidal attempts, as in \cite{greist1973computer}. 

In some other cases, however, the results on response quality were not fully conclusive, like in the early work by \cite{baker1992new}; moreover, the introduction of computer-mediated communication, especially in the last years when digital interactions have become pervasive, can lead to fatigue effects and negatively impact the collected information. For example, the effect of the survey technique on the respondent satisficing is explored by~\cite{heerwegh2008face}: they compare the results obtained in a face-to-face interview with those collected with a Web survey and they observe a higher degree of satisficing among the Web respondents, which would lead to lower data quality than that obtained in the face-to-face survey. 

In our work, we focus on Web-based communication and our research is aimed at finding a sweet spot between the convenience of a digital data collection and the risk of interacting with distracted respondents. Our approach aims to improve user acceptance and ``enjoyability'' of survey compilation as a means to an improved response quality.

\subsection{Chatbots, intelligent agents and conversational interfaces}
While the first example of chatbot, named ELIZA, was created by \cite{weizenbaum1966eliza}, this kind of applications resembling human interaction have been experiencing a second youth for the last few years. 

We can indeed distinguish several different kinds of those applications, as discussed by \cite{gao2019neural}: question-answering agents (single shot interaction, complex query), task-oriented dialogue agents (slot-filling or form-filling), chatbots (end to end conversation). Another classification of chatbots is offered by \cite{ramesh2017survey}.

Moreover, \cite{klopfenstein2017rise} propose a survey and history of conversational interfaces which leads to what they named ``botplication'', with its distinguishing features: thread-centric experience, history awareness, enhanced user interfaces, limited natural language processing capabilities, message consistency, guided conversations.

Those agents are also employed for diverse objectives. Without claiming to be exhaustive, some examples are as follows: \cite{angara2017foodie} present a smart kitchen assistant built on top of IBM Watson for recipe recommendation; \cite{gardiner2017engaging} introduce a conversational agent for awareness creation and behavioural change; \cite{panasiuk2018enabling} explain how to build domain-specific chatbots trained by mapping domain-specific knowledge to conversation/dialogue patterns; \cite{de2013aiml} focus on creating knowledge bases for chatbots with an educational purpose.

This new wave of applications led also to a renewed interest and investigation about their evaluation, assessment and limitations. Many authors explore the challenges of building an accurate training set for simulating credible question/answer conversation patterns: \cite{shah2018building} exploit crowdsourcing both to collect different formulations of relevant natural language expressions and to evaluate result quality.

On the other hand, building on their user experience expertise, \cite{badiu2018ux} claims that today's chatbots are far from being ``intelligent'', because they guide users through simple linear flows, and their user research shows that users have a hard time whenever they try to deviate from such flows. The authors discusses several characteristics required to improve user experience: the importance of pre-defined buttons and free-text user input, the possibility to follow non-linear flows, the importance of language and tone, the respect of user privacy. Moreover, user experience has been proved to be a weak point in chatbots, as explained by~\cite{luger2016like}, because conversational agents often fail to bridge the gap between user expectations and their actual operation capabilities.

Similarly, \cite{jain2018evaluating} present a study on first-time users of chatbots, with both a quantitative (time, messages) and qualitative (features) analysis. The authors propose a number of design implications: the need to clarify chatbot capabilities at start, the provision of content suitable for chat channel (i.e., minimizing user input), the possibility to maintain context, to keep a consistent chatbot ``personality'', to handle failures, to offer a mix of text/buttons/media.

Recently, initial studies on the adoption of intelligent chatbots in the administration of interviews and surveys have been performed. \cite{xiao2019tell} investigate the use of an AI-powered chatbot to conduct open-question surveys and they show a higher level of participant engagement as well as a better quality response w.r.t. a traditional survey. Research also focuses on the effects of chatbots in winning respondents trustworthiness, as in \cite{akbar2018effects}, in increasing self-disclosure, as in \cite{zhou2019trusting}, and in identifying users personality traits, as in \cite{zhou2019getting} and in \cite{xiao2019should}.

Even more advanced interaction modes are adopted to conduct interviews: virtual agents in the form of avatars are employed to increase self-disclosure and to create rapport by \cite{mell2017prestige}, and show also their benefits in the case of mental health reporting, as in \cite{lucas2017reporting}.

Overcoming the natural language capabilities of chatbots in interviewing is also investigated by \cite{zhou2019building} through a  human-assisted interviewing. \cite{kim2019comparing} conduct an experiment to explore the effect of platform and language style on response quality in closed-question surveys; in particular, by comparing a chatbot platform and a traditional Web survey, they show a potential advantage of the conversational approach to reduce satisficing and increasing data quality.

Our definition of ``conversational survey'' is slightly different than the one adopted in the above mentioned studies, in that we do not require any training of an intelligent chatbot system. Our research is oriented to build surveys in which the machine asks questions and the user replies, while chatbots usually adopt the opposite reactive pattern, with the intelligent agent interpreting user utterances and dynamically reacting to them. Our approach is more straightforward and similar to a traditional closed-question survey design; as explained in Section~\ref{sec:tool}, even if we allow for the insertion of open-questions with free-text answers, we do not provide any dynamic elaboration of user-generated text. 
Nonetheless, we leverage on the results of conversational interfaces research, because the questionnaire compilation experience in our tool is very similar from the user point of view, since it happens within a chat interface. In our design, we therefore give importance to simplify user interactions with effortless options (buttons and star-rating), we offer the possibility of designing different alternative branches, we focus on providing guided and consistent conversations. 

The rationale behind our design choices is motivated by the desire to fulfill several requirements: (1) assuring survey standardization, because the pre-defined conversation flow of our approach allows for repeatability and comparability of the collected answers, which can only be partially enforced by chatbots because of their reactive nature; (2) guarantee a coherent and possibly flowless user experience, which has been proved to be a weak point in chatbot UX; (3) making the survey design and setup similar to traditional survey, without the need for training a natural language understanding engine (which may require a large training set to create a machine learning model that leads to a meaningful user experience).

\subsection{Storytelling and user engagement}

The last ingredient we introduce in our conversational survey approach is storytelling, because we believe that the chat shape of our questionnaire allows for interleaving question/answer patterns with more colloquial interactions (cf. Section~\ref{sec:tool}) that can help in putting the survey into context. Indeed, attracting the attention of the user is of utmost importance for survey administration; a classification of user engagement approaches in the area of citizen participation is offered by \cite{celino2016towards}.

\cite{lambert2013digital} explains the basics of (digital) storytelling as a communication means. 
\cite{costikyan2000stories} believes that the continuum between stories and games led to the rise of game books in the '80s with alternative path choices.
Game books have been applied to different contexts, like education as in \cite{figueiredo2015development}, or health as in \cite{brandao2015augmented}, which use it in combination with augmented reality.

A popular tool to create interactive stories with a storytelling approach is Twine\footnote{Cf. \url{http://twinery.org/} (last visited: 2020/01/20).}, which can be used to design hypertexts or even games. \cite{friedhoff2013untangling} analyse the platform to highlight the focus on individual experience, the brainstorming-oriented user interface, and the openness of the platform itself. We indeed take Twine as an inspiration for our conversational survey editor (introduced in the next section).

In the area of questionnaire design, the only examples we found that shares some similarities with our approach are the one proposed by a company named Upinion, which highlights the market interest towards conversational survey in a white paper by \cite{ouass2018upinion} and the Surveybot platform\footnote{Cf. \url{https://surveybot.io/} (last visited: 2020/01/20).} which allows survey to be administered via Facebook, with the interaction limitations of the Facebook chatbot channel. The main providers of online survey tools currently focus on incremental improvements of user experience, by allowing the user to answer a question at a time in a sort of scrolling continuum that slightly resembles a chat. 

The toolkit we designed to create conversational surveys, therefore, on the one hand allows the questionnaire designers to build an interactive ``storytelling'', and on the other hand offers a natural chat interface to survey compilers to improve their experience and increase their engagement.

\section{Conversational Survey toolkit}\label{sec:tool}
On the basis of the considerations given in the previous section, and on our own experience in the design and development of user engagement tools, we came up with the concept of conversational survey, i.e. a method to administer questionnaires in a chat-like form, so that the compiler experiences it as if it was a conversation with another person rather than a pure survey. 

It is worth noting that our approach differs from those based on the adoption of chatbots and intelligent agents (like the mentioned works by \cite{xiao2019tell} and by \cite{kim2019comparing}), which imply some sort of natural language understanding (NLU) and artificial intelligence (AI). Our concept, on the contrary, is based on the idea that the survey is designed as a pre-defined conversation flow (with the possibility of branches, as explained in the following) that is experienced by the compiler through a chat interface. Avoiding the use of NLU, our goal is to overcome the limits in user experience when the system does not fully ``understand'' user's utterances, thus failing to meet user expectations, as illustrated by \cite{luger2016like}. As a consequence, our approach is more suitable for quantitative analysis than for qualitative research, even if the latter can be also partially addressed as illustrated in the following.

\begin{figure}[htb]
	\centering
		\includegraphics[width=.90\textwidth]{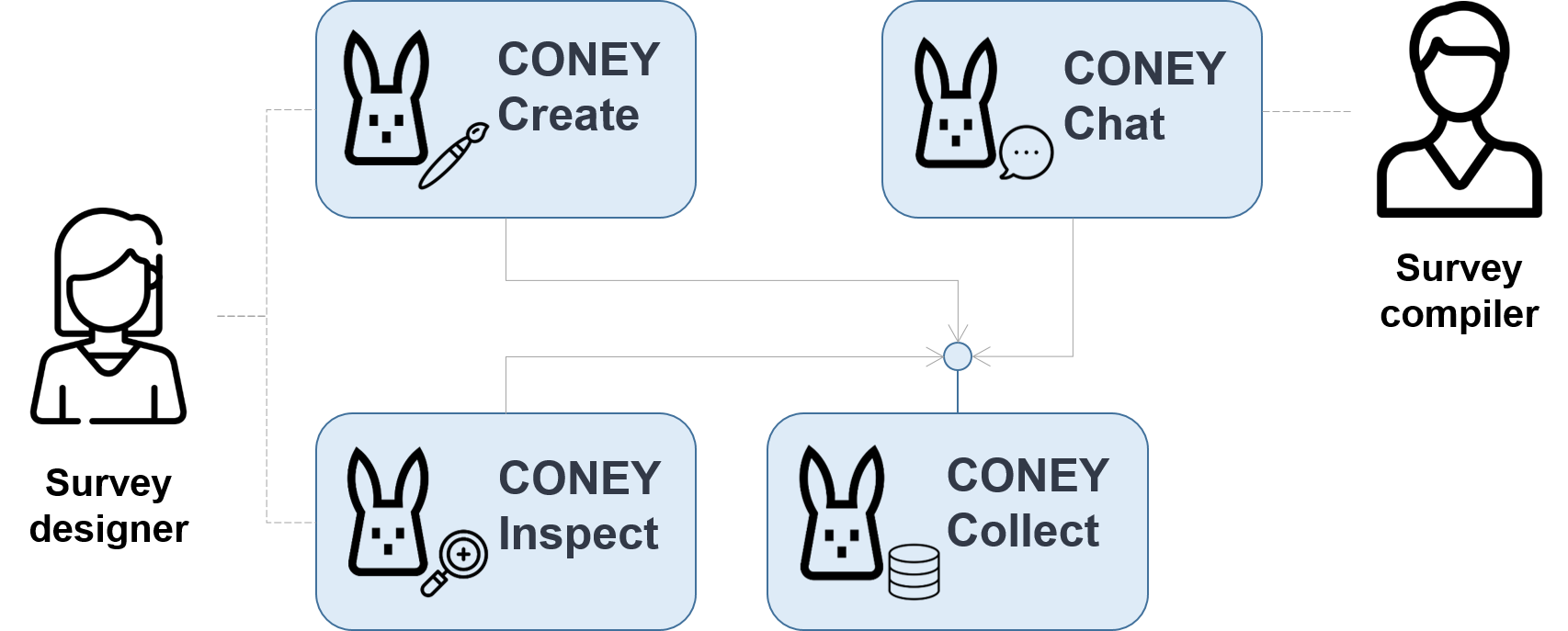}
	\caption{Components of the CONEY toolkit}
	\label{fig:coney-arch}
\end{figure}

We  designed and developed our toolkit named CONEY (CONversational survEY), which is composed of different components, as illustrated in Figure~\ref{fig:coney-arch} and explained hereafter.

\begin{figure}[htb]
	\centering
		\includegraphics[width=1.00\textwidth]{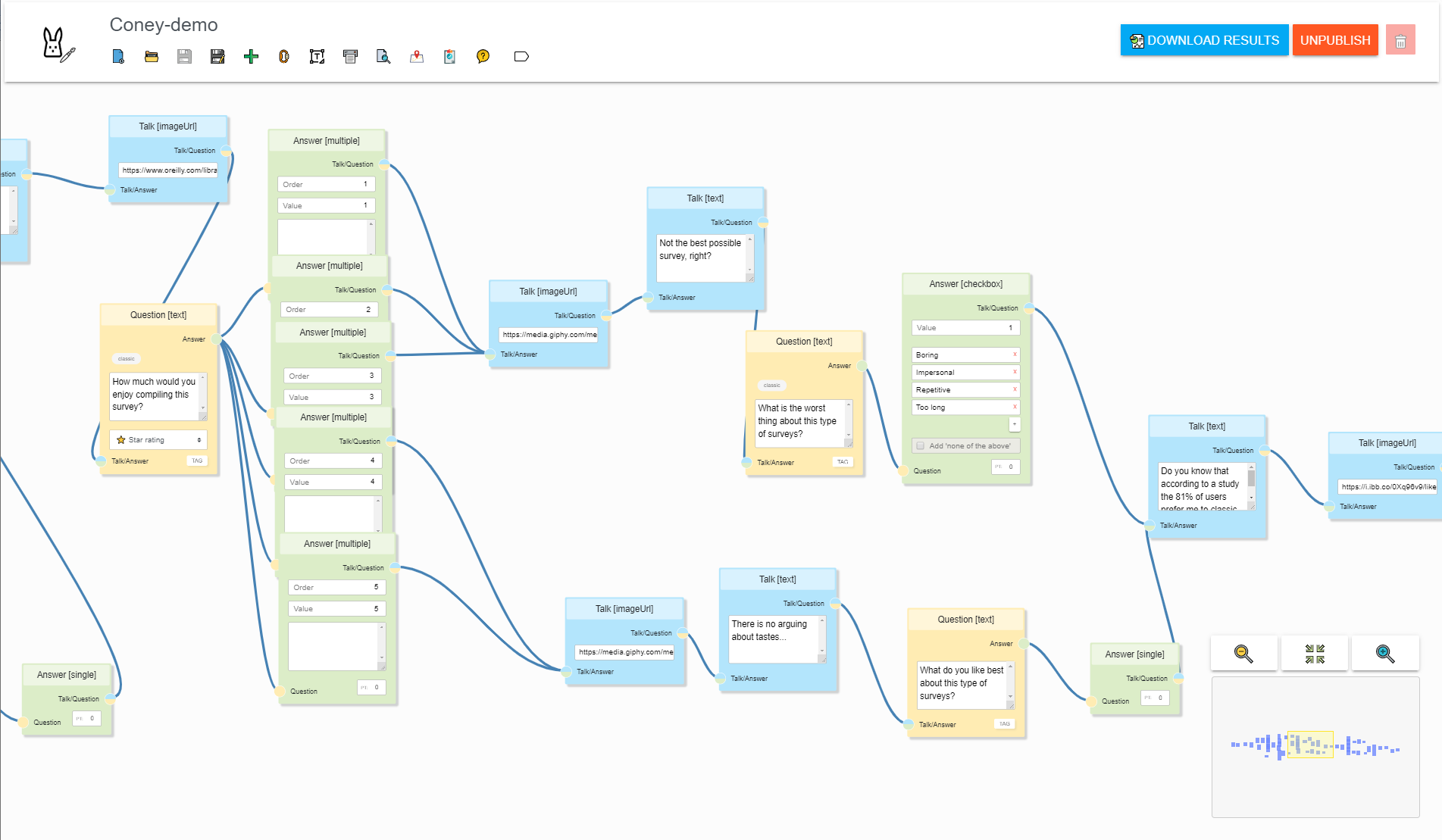}
	\caption{Screenshot of the CONEY Create editor to design conversational surveys; each block represents a question, an answer or a ``colloquial'' element like text, image or hyperlink; the editor supports the designer in creating and connecting the building blocks, as well as in enabling alternative paths in the conversation; the ``preview'' button on top allows to test the interaction flow before publishing.}
	\label{fig:coney-create}
\end{figure}

\textbf{CONEY Create} (cf. Figure~\ref{fig:coney-create}) is the graphical editor for the survey designer to create the questionnaire in the form of a conversation; inspired by \cite{friedhoff2013untangling}, the editor is a drag-and-drop tool which allows for the design of a questionnaire in a ``hypertextual'' fashion with text, question and answer blocks (respectively, blue, yellow and green boxes in the picture) and the possibility to create alternative branches depending on the compiler choice (i.e., depending on the chosen answer, the conversation flow continues in different ways, for example to ask clarification questions).
 
The editor offers plenty of question types: single choice questions with different answer visualizations (buttons, star-rating, emoticons, slider), multiple choice questions (with check-box answers) and open question; as explained above, since there is no use of AI, the open questions allow for free-text answers, but no elaboration of the provided text is made: compilers' answers are only collected for post-hoc analysis. The ``conversation flow'' approach allows for a storytelling, enhanced by the possibility to include colloquial and multimedia content.

Finally, question blocks can be annotated with a label to indicate the respective investigated latent variable, while answers can be annotated with the respective numerical coding: this kind of information is reused at answer analysis time, as explained in the CONEY Inspect component.

The editor itself does not constrain the survey designer to the use of a specific language style, and she has the responsibility also for the ``storytelling'' design. However, the tool allows for saving and reusing question/answer patterns across different surveys, as well as for cloning an existing survey to adapt it to a different usage scenario.

CONEY Create is developed with basic Web technologies,  with the Angular framework\footnote{Cf. \url{https://angular.io/} (last visited: 2020/01/20).} and by leveraging the Rete.js framework for visual programming\footnote{Cf. \url{https://rete.js.org/} (last visited: 2020/01/20).}.
 
\begin{figure}[htb]
	\centering
		\includegraphics[width=1.00\textwidth]{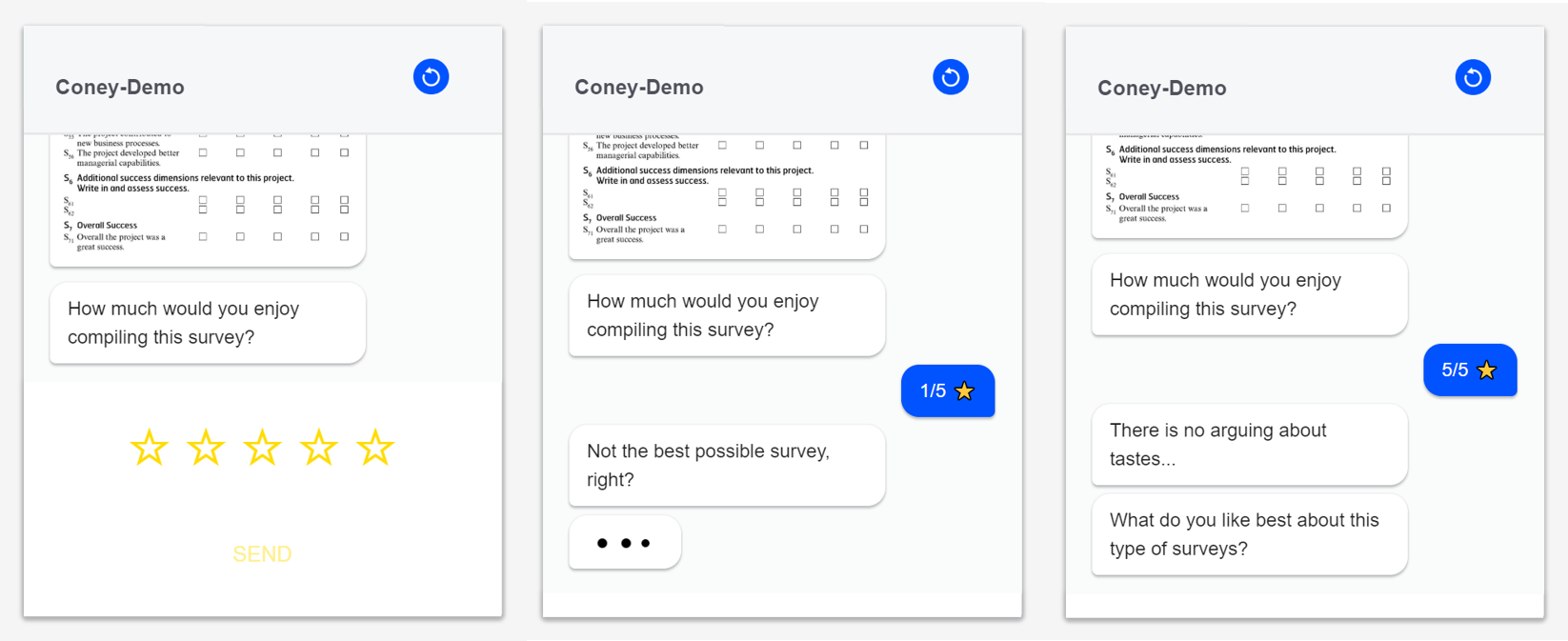}
	\caption{Screenshots of the CONEY Chat user interface for survey compilers; the questionnaire appears as a dynamic chat and the flow of messages depends on the previous answer choice (cf. alternative branches in Figure~\ref{fig:coney-create}). }
	\label{fig:coney-chat}
\end{figure}

\textbf{CONEY Chat} (cf. Figure~\ref{fig:coney-chat}) is the Web-based user interface to administer the designed survey to compilers in the form of a chat; in this case, the inspiration comes from chat clients and popular mobile apps like Whatsapp, Messenger or Telegram. The user experiences a seamless flow thanks to the personalized path based on his/her answers. Furthermore, even when the survey is a purely quantitative research method (with closed questions with numerically-coded answers), the interaction style makes it resemble an interview, i.e. a qualitative research approach.

The interested reader can try CONEY Chat by experiencing our demo survey at \url{http://bit.ly/try-coney}.

Also CONEY Chat is a Web application developed with the Angular framework and with responsive design for an optimal experience on any device, including mobile phones. We are currently exploring the possibility to extend the toolkit to enable the survey administration through the most common social applications (e.g. Telegram), even if this would restrict the rich interaction features offered by our tools.

\textbf{CONEY Collect} is the back-end component that gathers and store the data related to both the survey design and the survey responses. The storage makes use of the Neo4J\footnote{Cf. \url{https://neo4j.com/} (last visited: 2020/01/20).} graph database, which was selected as the most suitable data management approach, because the conversation flow of the surveys is indeed well represented by a graph structure; this component offers a Web API, developed with the Spring Data framework for Neo4J\footnote{Cf. \url{https://spring.io/projects/spring-data-neo4j} (last visited: 2020/01/20).}, which provides CRUD operations to the other toolkit components.

\begin{figure}[htb]
	\centering
		\includegraphics[width=1.00\textwidth]{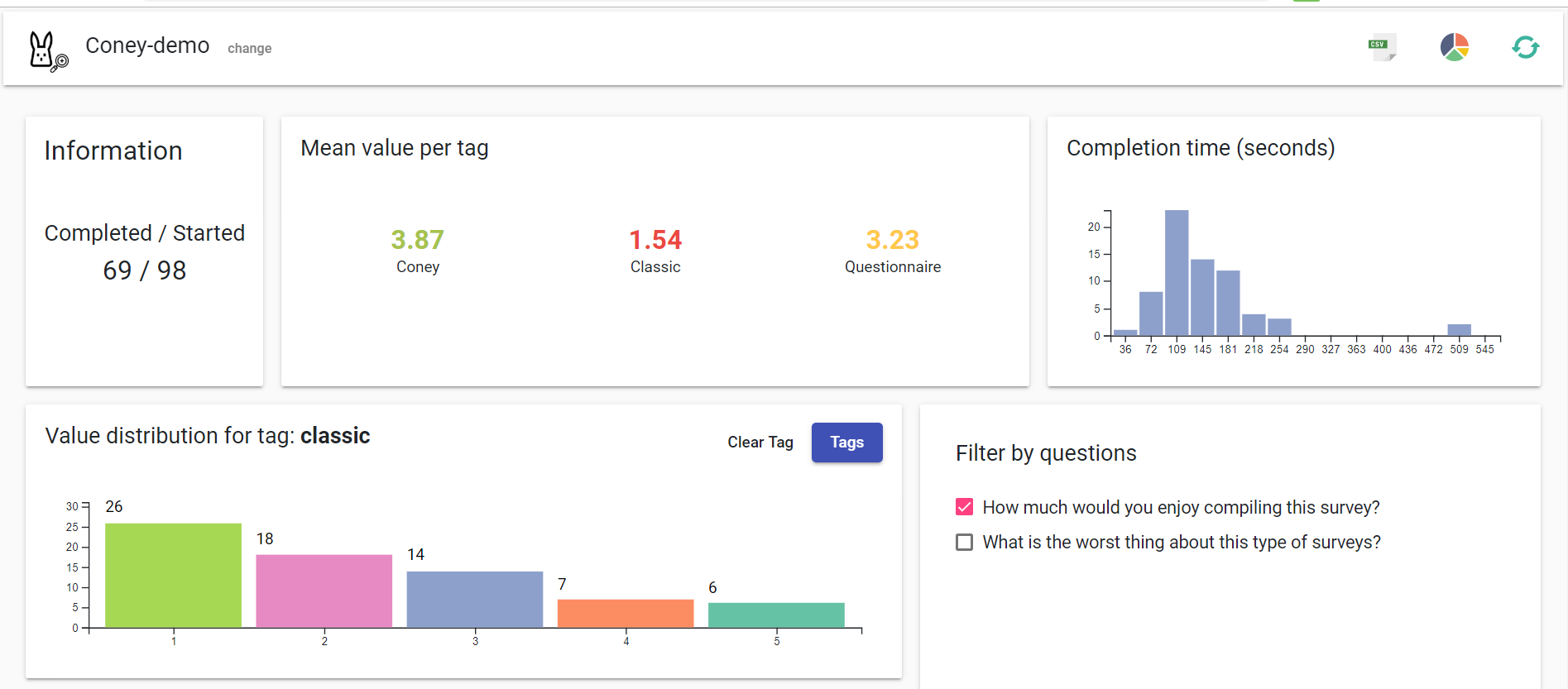}
	\caption{Screenshot of the CONEY Inspect dashboard with basic result statistics}
	\label{fig:coney-inspect}
\end{figure}

Finally, the \textbf{CONEY Inspect} component (cf. Figure~\ref{fig:coney-inspect}) is the dashboard application for the survey analyst to simplify the statistical analysis of the answers collected through the conversational survey. The dashboard offers basic indicators, like the number of started and completed surveys, the average values for latent variable, the distribution of compilation time and the histograms of the answers per question; for more detailed analysis, the dashboard allows for the download of all collected answers in the form of a CSV file.
 
We would like to stress that we focus on the user experience, so that our goal is to provide a new and innovative interaction pattern from the compiler point of view; this of course implies a change in the survey design process, which not only is oriented to select and formulate the questions to collect relevant data, but should also take into account the different interaction pattern in the overall formulation of the questionnaire. We recognize that this represents an extra-effort for the survey designer, which is still to be fully tested and investigated; in this paper, we focus on the user experience and on the effectiveness of data collection.

\section{Experimental setup}\label{sec:exp}
To evaluate the ability of a conversational survey to act as a valid quantitative research tool, we designed several experiments. We aim to compare and contrast the results obtained through the submission of a traditional quantitative survey and a conversational one, i.e. administered through a UI like the one displayed in Figure~\ref{fig:coney-chat}. On the one hand, we aim to compare the user appreciation and acceptance with respect to the two methods and, on the other hand, we aim to investigate the feasibility of adopting the conversational method as a valid substitute of a traditional approach. In the rest of this section, we describe the two sets of experiments we performed to evaluate the two different aspects. In the following sections \ref{sec:ux} and \ref{sec:meth}, we will discuss the obtained results.

\subsection{The mobile banking questionnaire: survey content}\label{sub:content}
For our first set of experiments, we selected a questionnaire from the literature, which investigates the factors influencing the adoption of mobile banking solutions (\cite{kim2009understanding}). It is a traditional quantitative method, consisting of a list of 21 questions (investigating 7 latent variables); for each question, the user is asked to judge his/her level of agreement on a Likert scale. The full questionnaire is reported in \ref{app:traditional}. The traditional survey was proposed to users through the popular SurveyMonkey platform.

We created a conversational version of the mobile banking questionnaire, which consists of a chat dialogue, in which the user can answer to questions by selecting the most suitable options between a set of predefined alternatives, shown as buttons or star-rating scales in the chat interface. The conversational survey includes 7 questions (one per each of the latent variables of the original questionnaire) and some more colloquial contents to give the user the impression of an actual chat. The full dialogue is presented in \ref{app:conversation}. The conversational survey was presented to users through a custom Web interface developed by us.

Additionally, at the end of the compilation of the mobile banking questionnaire, each survey compiler was asked to answer a set of additional questions to evaluate the user appreciation and acceptance. Those questions consisted in 6 statements about the questionnaire compilation experience (``I found it interesting'', ``It was too long'', ``I enjoyed compiling it'', ``It was boring'', ``I found it intuitive to answer its questions'', ``I was a bit confused in answering its questions'') and the compiler was asked to rate his/her level of agreement/disagreement on a 5-level Likert scale. The user experience part of the survey was always proposed through SurveyMonkey. A final optional question asked the compilers to provide any free-text feedback they deemed relevant.

\subsection{The mobile banking questionnaire: data collection}\label{sub:crowds}
We collected the answers to our traditional and conversational surveys, as well as for the user experience questions, in different experimental settings. An  A/B testing  was implemented by two parallel campaigns: a first ``traditional'' campaign (T campaign), completely run on SurveyMonkey, in which the participants compiled the original questionnaire and then answered the 6 user acceptance questions; a second ``conversational'' campaign (C campaign), in which the participants compiled the chat-version of the questionnaire through the dedicated Web interface, and then were redirected to SurveyMonkey where they answered the 6 user acceptance questions. The results of the A/B testing are useful to compare the unbiased opinion of people with respect to the two alternatives.

A second experimental setting was aimed to a direct comparison of the two tools: the participant compiled both the traditional and the conversational versions of the questionnaire, then answered to the 6 user experience questions for each of the two versions, and finally were asked to express their explicit preference between the two survey methods. In order to avoid introducing ordering bias, we run two campaigns: a ``first traditional then conversational'' campaign (TC campaign) and a ``first conversational then traditional'' campaign (CT campaign), in which the order of compilation of the two surveys changed; of course, the user acceptance questions were fed after the compilation of both surveys.

The participants changed in each of the four campaigns, i.e. different people participated to the T, C, TC and CT campaigns. We recruited our users through the Prolific crowdsourcing platform (\cite{palan2018prolific}). In contrast to other crowdsourcing platforms, which are mostly oriented to atomic task execution, Prolific is designed for researchers and their needs and specifically suitable for social and economic science experiments (\cite{peer2017beyond}); moreover, the ``workers'' panel is quite varied and more than half of the participants have a bachelor degree or higher\footnote{Cf. \url{https://www.prolific.ac/demographics/} (last visited: 2020/01/20).}. 

The participants were briefed with the request of testing the formulation of a mobile banking questionnaire. They were asked to answer sincerely, in order to collect relevant information, but they were also asked to evaluate the questionnaire itself, in order to provide us feedback on how to possibly improve it. Nothing was explicitly said in the brief about the survey methods or tools.

For each of the four campaigns (T, C, TC and CT campaigns), we recruited 100 participants; the participants changed in each campaign, therefore in total we involved 400 distinct people. The profiles of the user panel is as follows: between 18 and 60 years old (mean 34, median 32), 40\% male and 60\% female, 83\% from UK and 17\% from Italy, 69\% of participant with a part-time or full-time job and only 22\% students. All the experiments were run in November-December 2018.

\subsection{The motivation to participate questionnaire: survey content}\label{sub:content2}
For our second set of experiments, we selected a questionnaire from \cite{levontin2018motivation}, which extends the \cite{schwartz2005basic}  model of basic values to investigate the motivations influencing the participation of people to collective actions; we customized that questionnaire to apply our investigation to the motivations of crowdsourcing platform participants (the so-called mcro-workers). We selected a set of 10 latent variables and 2 questionnaire items for each of them; additionally, we inserted a final item to measure the global level of motivation.

We prepared 3 versions of the motivation questionnaire: a formal formulation, in which all items are expressed as statements to be evaluated on a 1-5 Likert scale, which was implemented both on a traditional survey platform (SurveyMonkey) and on our conversational toolkit; an informal formulation, in which the same items are completely re-phrased and have different response options (buttons, star-rating, slider, etc.), which was implemented only on Coney. The different formulations are reported in \ref{app:formal} and \ref{app:informal}. 

\subsection{The motivation to participate questionnaire: data collection}\label{sub:crowds2}
As in the first set of experiments, we performed an A/B testing by launching a crowdsourcing campaign on Prolific for each of the 3 questionnaire versions. The participants were briefed with the request of providing their honest feedback about the motivations behind their own participation to the Prolific crowdsourcing platform. Nothing was explicitly said in the brief about the survey methods or tools.

For each of the 3 campaigns (traditional formal TF, conversational formal CF and conversational informal CI campaigns), we recruited 100 participants; the participants changed in each campaign, therefore in total we involved 300 distinct people. The profiles of the user panel is as follows: between 18 and 66 years old (mean 32, median 29); 45\% male and 55\% female; 42\% from UK, 17\% from Portugal, 16\% from Poland, 8\% from Italy and the remaining 17\% from other 11 EU countries (Greece, Spain, Germany, Finland, France, Ireland, Norway, Denmark, Netherlands, Slovenia, Sweden); 67\% of participant with a part-time or full-time job and only 30\% students. 
All the experiments were run in October 2019. 

\section{User experience evaluation}\label{sec:ux}
The first hypothesis we want to test is whether it is true that users prefer a conversational interface to answer a questionnaire rather than a traditional method. The research question can be formulated as follows: 
\begin{quote}
	[RQ1] \emph{Do users prefer a conversational survey to a traditional survey? If they do, what factors influence such preference?}
\end{quote}
In the rest of this section, we illustrate the experiments we executed to answer this question. We specifically refer to the mobile banking questionnaires described in the previous \S~\ref{sub:content} and \ref{sub:crowds}.

\subsection{A/B testing between subjects}\label{sub:ab-test}
The first experiment to answer RQ1 is the A/B testing through the T and C campaigns: one (control) group was given the traditional version of the questionnaire, while one (treatment) group was asked to answer the conversational version of the survey. Both groups were asked to answer a set of final questions on user experience.

We compared the results obtained in the 6 user acceptance questions, with a two-sample statistical hypothesis testing for the difference in mean. The null hypothesis is that the mean values of each answer is the same across the two groups, while the (two-tailed) alternative hypothesis is that the mean value is different between the two groups. Since each question asked for a 1-5 Likert-scale agreement answer, we compared the numerical results obtained from the two groups of respondents; given the non-normal distribution of the collected data points, the t-test cannot be applied, therefore we adopted the Wilcoxon signed-rank unpaired test (also known as Wilcoxon T test). The results illustrated in Table~\ref{tab:ux-wilcox} show that a statistically significant difference is indeed recorded: users find the conversational survey more interesting and intuitive (+10\%), more enjoyable (+5\%) and less boring (-18\%); there is no difference regarding the potential confusion (low value for both surveys), while they found the conversational version slightly longer: indeed it was the case that the average time to complete the conversational survey was higher than the one to compile the traditional questionnaire.

\begin{table}[htb]
\begin{tabular}{lcccc}
\toprule
            & Traditional    & Conversational  & Wilcoxon test                      & Mean       \\ 
            & mean (std dev) & mean (std dev)  & p-value                            & difference \\ 
\midrule                                                                                       
Interesting & 3.48 (0.87)    & 3.84 (0.92)     & 0.002 **\textcolor{white}{*}       & + 10 \%    \\ 
Intuitive   & 3.69 (0.77)    & 4.11 (0.79)     & 0.000 ***                          & + 10 \%    \\ 
Enjoyable   & 3.49 (0.67)    & 3.68 (0.97)     & 0.028 *\textcolor{white}{**}       & + 5 \%     \\ 
Boring      & 2.49 (0.94)    & 2.11 (0.94)     & 0.003 **\textcolor{white}{*}       & – 18 \%    \\ 
Confusing   & 1.73 (0.81)    & 1.68 (0.83)     & 0.560 \textcolor{white}{***}       & – 3 \%     \\ 
Long        & 1.57 (0.56)    & 1.78 (0.69)     & 0.032 *\textcolor{white}{**}       & + 12 \%     \\
\bottomrule
\end{tabular}
\caption{User acceptance by A/B testing: difference in mean between conversational and traditional survey (N=100 for each survey; *** p$<$0.001, ** p$<$0.01, * p$<$0.05).}
\label{tab:ux-wilcox}
\end{table}

\subsection{Direct comparison within subjects}\label{sub:comparison}
To further investigate the relationship between the different aspects of user experience, we performed a second experiment. This time, we asked participants to compile both the traditional and the conversational questionnaires; after completing the surveys, users were again asked to judge their experience using the 6 dimensions introduced in the A/B testing setting, this time for each of the survey separately. 

As introduced in Section~\ref{sub:crowds}, a total of 200 respondents were involved through the Prolific platform; half of them compiled the traditional survey first and the conversational survey after (TC campaign); the other half compiled the questionnaires in the reverse order (CT campaign). Indeed, a closer analysis of the collected answers show that the order of compilation had no significant effect of the results: a multivariate analysis of variance for the effect of order did not yield a statistically significant difference between the two conditions (Pillai test statistics 0.155, p-value 0.281); the ANOVA tests for the effect of order on individual variables always rejected the null hypothesis at the 10\% level. 

The same Wilcoxon T-test used in the A/B testing was performed (this time as a paired test) over the 200 answers collected in the CT/TC campaigns for each of the 6 user experience dimensions. Again, the null hypothesis is that there is no difference in means between the experience on the traditional questionnaire and on the conversational survey, the two-tailed alternative hypothesis is that such difference exists. The results are displayed in Table~\ref{tab:ux-wilcox-2}. 

In this case, the difference in mean is statistically significant for all 6 dimensions and such difference is larger and always showing an advantage for the conversational survey method. In other words, the participants think that the conversational survey is more interesting, more intuitive, more enjoyable, less boring, less confusing and less long than the traditional questionnaire. 

\begin{table}[htb]
\begin{tabular}{lcccc}
\toprule
            & Traditional    & Conversational  & Wilcoxon test                      & Mean       \\ 
            & mean (std dev) & mean (std dev)  & p-value                            & difference \\ 
\midrule                                                                                       
Interesting & 3.19 (0.88)    & 4.19 (0.74)     & 0.000 ***                          & + 31 \%    \\
Intuitive   & 3.44 (0.91)    & 3.90 (0.81)     & 0.000 ***                          & + 13 \%    \\
Enjoyable   & 3.10 (0.86)    & 4.01 (0.80)     & 0.000 ***                          & + 29 \%    \\
Boring      & 3.02 (1.06)    & 1.80 (0.71)     & 0.000 ***                          & – 40 \%    \\
Confusing   & 2.00 (0.89)    & 1.77 (0.79)     & 0.004 **\textcolor{white}{*}       & – 12 \%    \\
Long        & 2.54 (0.98)    & 1.83 (0.70)     & 0.000 ***                          & – 28 \%    \\
\bottomrule
\end{tabular}
\caption{User acceptance by direct comparison: difference in mean between conversational and traditional survey (N=200; *** p$<$0.001, ** p$<$0.01).}
\label{tab:ux-wilcox-2}
\end{table}

It is curious to notice that, even if the measured compilation time was again slightly longer on average for the chat-like interface, the users got a different impression from the direct comparison and more strongly penalized the traditional agree-disagree questionnaire. It is true that the number of questions in the two versions is indeed different (7 in the chat and 21 in the agree/disagree questionnaire), but the colloquial approach, with the chat messages appearing one at a time as in a real conversation, forces the users to take more time in answering the questions; this is a designed feature of our conversational survey tool, to cope with the risk of compilers going very fast through the Likert scale questions of the traditional version, without actually reflecting on their answers. This is also testified by the fact that some recent commercial survey tools (like the already mentioned TypeForm) indeed introduce tweaks in the user interface to force the user to focus on and think to one question at a time. 

We can conclude that our experimental data clearly show that a conversational survey tool is ``better'' perceived and accepted than a traditional survey method.

\subsection{Survey method preference}\label{sub:pref}
Additionally, in the direct comparison experiment, all users were asked to explicitly express their preference between the two questionnaire versions, with a single question ``Which version of the survey do you prefer?'' with 5 possible answer options (``I definitely prefer version 1 ``, ``I partially prefer version 1'', ``The two versions are the same to me'', ``I partially prefer version 2'', ``I definitely prefer version 2'').

\begin{figure}[tb]
	\centering
		\includegraphics[width=.70\textwidth]{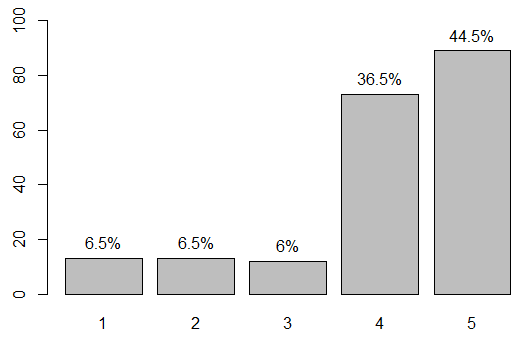}
	\caption{Distribution of Likert-scale answers regarding the direct preference between the two types of questionnaires, where 1 means strong preference for the traditional agree-disagree questionnaire, while 5 means strong preference for the chat-like version.}
	\label{fig:preference}
\end{figure}

The results on the direct preference question show a clear majority of respondents (\smalltilde81\%) that partially or strongly favour the conversational survey; Figure~\ref{fig:preference} displays the distribution of collected answers. 

Furthermore, we conducted a Chi square test to evaluate if any significant relation exists between preference and population characteristics (cf. Table~\ref{tab:chisquare}). We measured no significant effect of gender, employment status, student status, nationality or native language on the survey method preference. On the other hand, there is a significant effect of age on the preference; still further investigation does not show a significant correlation (non-significant Kendall tau), so we cannot conclude that the preference is stronger for younger or older participants.

\begin{table}[htb]
	\centering
\begin{tabular}{lc}
\toprule
Population characteristics          & Chi square p-value                \\
\midrule  
Gender               & 0.264 \textcolor{white}{*} \\
Employment status    & 0.606 \textcolor{white}{*} \\
Student status       & 0.699 \textcolor{white}{*} \\
Nationality          & 0.610 \textcolor{white}{*} \\
Native language      & 0.549 \textcolor{white}{*} \\
Age                  & 0.011 *                    \\
\bottomrule
\end{tabular}
\caption{Significance of the effect of population characteristics over survey preference with Chi square test (N=200; * p$<$0.05). No feature has a significant effect apart from age, which however has non significant  correlation with preference (Kendall's rank correlation p-value=0.274).}
\label{tab:chisquare}
\end{table}

\subsection{Factors influence}\label{sub:cfa-ux}
In the direct comparison experiment, we also performed a multiple regression statistic test, to estimate the possible interplay between the measured variables. We chose, as independent variables of the model, the difference between the scores on the conversational and traditional methods received on the 6 user experience dimensions and, as dependent variable, the answer on the preference question. The hypotheses we want to test are as follows: an increase in interestingness, enjoyability and intuitiveness and a decrease in length, boringness and unintelligibility have a positive influence on the preference of the conversational method.

\begin{figure}[hbt]
	\centering
		\includegraphics[width=.75\textwidth]{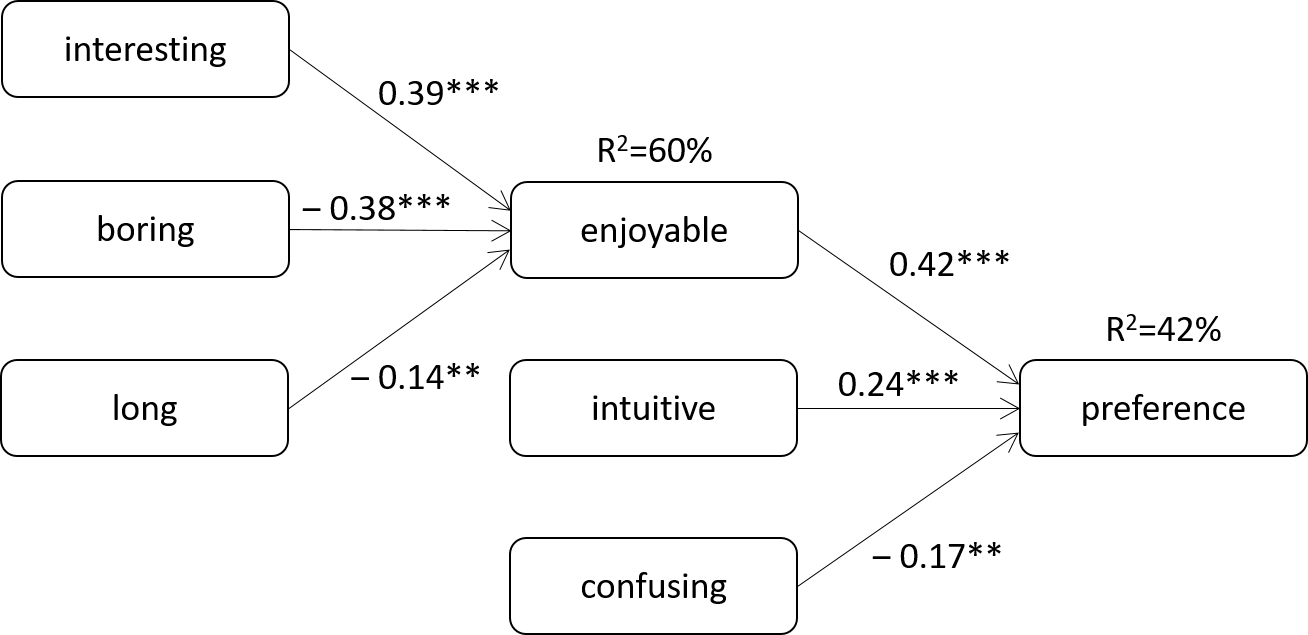}
	\caption{Factors influence on method preference (N=200; *** p$<$0.001, ** p$<$0.01)}
	\label{fig:cfa-ux}
\end{figure}

After a first exploratory model in which all independent variables influence the method preference, through a confirmatory factor analysis we came up with a hierarchical model that well explains the relation between the variables. The resulting model, depicted  in Figure~\ref{fig:cfa-ux}, confirms our hypotheses and shows very positive goodness of fit metrics (CFI=0.987, TLI=0.971, RMSEA=0.062, SRMR=0.025). 

This analysis shows that the preference granted to the conversational method is indeed influenced by a judgment of higher enjoyability and intuitiveness and a lower unintelligibility; moreover, our participants think that the approach enjoyability is in turn dependent on higher interestingness and lower boringness and length.

\subsection{Qualitative analysis}\label{sub:comments}
In all crowdsourcing campaigns conducted on the Prolific platform -- both during the A/B testing and during the direct comparison campaigns -- the participants were given the option to add some free text comment on any aspect of the study. While those that left and actual textual feedback were a minority of participants (around 18\%), it is interesting to note both the content and the conditions under which such observations were provided. Indeed, in the campaigns' brief, we only told the participants that we were putting together a mobile banking questionnaire and that we were testing different question formulations.

In the A/B testing campaigns, those who compiled the traditional survey only almost did not provide any feedback (around 14\% of T campaign participants wrote something in the feedback box, mostly saying ``thanks'' or ``nothing to comment''); only a couple of respondents asserted that a specific word used in the questionnaire was potentially misleading; no feedback was given on the questionnaire experience. On the contrary, those who compiled only the conversational survey (around 20\% of C campaign participants) and those who compiled both (around 21\% of CT and TC campaign participants) were more generous and ``talkative'' and left a number of interesting comments on the conversational tool itself. 

In general, the comments were very positive: ``I quite liked the lay out of the chat questions, i found it easier to follow'', ``I enjoyed the text message format, it was unique'', ``Great concept. Enjoyed the less robotic feel of this survey'', ``Answering a questionnaire through a chat is a cool idea to keep attention'', ``Really enjoyed this survey I especially enjoyed the format in which the survey was presented'', ``I really liked that the study resembled a chatroom, it's definitely unique and never seen before. I enjoyed this quite a lot!'', ``Great survey, loved the interaction'', ``I complete a lot of surveys and the chat one was much more interesting!''.

One of the participants to the C campaign also provided quite an extensive reflection on the experience and the potential advantages of the conversational survey:
\begin{quote}
\emph{A very novel way of answering a survey, it certainly held my attention better than a regular check box style survey. I would suspect as it was formed more like a message exchange I felt I should engage more and I wasn't able to have a predetermined answer ready as I didn't know what question was next. I liked it.}
\end{quote}
The feedback above indeed illustrates that, with respect to a traditional quantitative survey method, in which users quickly go through questions that appear in the same format, the timed interaction of the chat forces the user to wait and read each message exchange, possibly bringing a higher level of attention during the compilation.

The few critical comments to the conversational approach came from both those who compiled only the chat and those who answered both survey versions; most criticisms are related to the language used in the survey (``I disliked the language used in the chat version -- it felt too informal and a bit patronising'', ``some of the auto responses were a bit cheesy and annoying when you know it's not a real person'', ``perhaps less of the snarky replies please'', ``The chat was very condescending and unbelievable. I hate it when an obviously automated script tries to appear amusing and/or human'', ``The fake friendly nature of the chat version put me off''), even if it also seems that language can be quite a subjective topic, since some users liked it (``It was funny to see some of the autoresponses. I'd much rather carry out surveys in this format'', ``Really enjoyed the style of the texting version!''). 

Those comments raise an important issue in relation with the conversational survey approach: it is critical to choose the right language and tone. While this is a common issue in any survey design methodology (cf. for example \cite{saris2014design}), it is even more important in this case, since the chat-like channel resembles an actual dialogue with another person, much more than in the case of an ``aseptic'' questionnaire. Though, in the case of our experiments, some participants thought that we may have exaggerated with a colloquial style. An important research investigation, therefore, should be oriented to understand how to balance the dialogue style of a conversation with the correct and effective formulation of the survey questions. 

Some other suggestions we collected from the participants, especially from those who compiled both surveys, regarded the level of details of the questionnaire (``Perhaps a more tailored response and a greater exploration of each answer'', ``I was expecting much more specific questions'', ``I prefered the chat but I feel as if the options were more limited'') and the compilation time (``The text loaded quite quickly on the chat version so I felt a little rushed to read it'', ``I think you underestimate how long it takes to read the questions'', ``When I compile a survey, I want to do it in the fastest way. The chat is a nice thing but is also too slow''). Therefore, proper attention should be also given to a correct balance of the question specificity and the overall length.

\subsection{Conclusions on user experience}\label{sub:ux-concl}
Our experimental results clearly demonstrate that users indeed have a preference on the conversational survey with respect to a traditional questionnaire; in particular, users appreciated the ``enjoyability'' of the conversational tool and found it more interesting, more intuitive and less boring than its traditional counterpart. The novelty of the tool may indeed had influenced the results in favour of the chat-like system; still, there is a limited set of respondents ($<$13\%) who prefer traditional survey methods. A clear point of attention is the language adopted in the conversational survey: while a more colloquial wording may help in  putting the user at ease, a balance should be found to avoid the risk of making the survey too informal. 

\section{Method effectiveness evaluation}\label{sec:meth}
The second hypothesis we want to test is whether it is true that compiling a conversational questionnaire is comparable to adopting a traditional survey method. The research question can be formulated as follows:
\begin{quote}
	[RQ2] \emph{Is a conversational survey functionally equivalent to a traditional survey? Can the conversational approach be at least as reliable as a classical quantitative methods? Can the response quality of the conversational approach be at least as good as the one of a traditional method?}
\end{quote}
In the rest of this section, we illustrate the analyses performed to answer this question. We specifically refer to the data collected in the survey described in \S~\ref{sub:content2} and \ref{sub:crowds2}.

To test the conversational approach reliability, we perform a comparative evaluation based on a measure of internal consistency (Cronbach alpha) and on a measure of equivalence based on inter-rater reliability (Krippendorff alpha). To test the response data quality, we perform a comparative evaluation of satisficing based on the differentiation index between the given answers.

\subsection{Method reliability: item consistency in the surveys}\label{sub:cronbach}
One of the most widely-used reliability measures of surveys and questionnaires is Cronbach's alpha, which is a metrics of internal consistency of a test (\cite{cronbach1951coefficient}). By computing the metrics over a set of responses to different items, it is possible to evaluate whether the considered items measure the same phenomenon.

First of all, we computed  Cronbach's alpha for each of the three versions of the questionnaire. The results are reported in Table~\ref{tab:cronbach-global}: Cronbach's alpha is very similar for all the surveys; the \cite{feldt1987statistical} test of significance for the difference between alpha coefficients fails to reject the null hypothesis at the 10\% level in all three pairwise comparisons. 

\begin{table}[htb]
\begin{center}
\begin{tabular}{lccc}
\toprule
                & Traditional  & Conversational  & Conversational \\ 
                & Formal       & Formal          & Informal       \\ 
\midrule                                                             
Cronbach alpha  & 0.829        & 0.835           & 0.885          \\
\bottomrule      
\end{tabular}
\end{center}
\caption{Cronbach alpha coefficients measured for the three questionnaires separately (N=100 for each survey).}
\label{tab:cronbach-global}
\end{table}

On this first analysis, we can conclude that all questionnaires show a very good internal consistency\footnote{While there is no threshold to tell apart consistent and inconsistent coefficients, the commonly accepted rule-of-thumb is that a good level of consistency is indicated by alpha values above 0.7, and all the three versions have a value grater than 0.8.} and the coefficients do not show a statistically significant difference.

\subsection{Method reliability: equivalence of the surveys}\label{sub:krippen}
In order to test for the equivalence of the different study formulations, we measured the inter-rater reliability, i.e. the degree of agreement between the survey respondents. In our case, we do not aim to reach a high value of agreement, but to test whether the respondents show the same tendency to agree in the three studies.

For this evaluation we computed the alpha coefficient defined by \cite{krippendorff1970estimating} for each survey version (TF, CF and CI); the results are displayed in Table~\ref{tab:krippen}. 

\begin{table}[htb]
\begin{center}
\begin{tabular}{lccc}
\toprule
                   & Traditional\textcolor{white}{***} & Conversational\textcolor{white}{***} & Conversational\textcolor{white}{**} \\
                   & Formal\textcolor{white}{***}      & Formal\textcolor{white}{***}         & Informal\textcolor{white}{**}       \\
\midrule                                                             
agreement          & 0.343 \textcolor{white}{***}      & 0.363 \textcolor{white}{***}         & 0.323 \textcolor{white}{**}         \\
alpha & 0.109 \textcolor{white}{***}   & 0.105 \textcolor{white}{***}         & 0.129 \textcolor{white}{**}         \\
p-value            & 0.000 ***   & 0.000 ***      & 0.003 **       \\
\bottomrule      
\end{tabular}
\end{center}
\caption{Krippendorff alpha coefficients measured for the three questionnaires separately (N=100 for each survey; *** p$<$0.001, ** p$<$0.01).}
\label{tab:krippen}
\end{table}

Then, we tested the difference between the alphas by using the approach suggested in \cite{gwet2016testing}; the result is that all three pairwise comparisons (TF-CF, CF-CI, TF-CI) are not statistically significant at the 10\% level, thus the null hypothesis of difference between the alphas is rejected. This means that, from a method reliability point of view, the three studies are equivalent despite their differences in both adopted tool and formulation; we can conclude that our conversational approach is at least as reliable as a traditional method, in terms of inter-rater reliability.

\subsection{Method response quality: satisficing effect}\label{sub:satisficing}
Satisficing is a decision-making strategy that people adopt to choose an acceptable alternative that reduces the cognitive workload associated with performing an activity. With respect to questionnaires, satisficing appears in various forms, as discussed by \cite{krosnick1991response}: choosing a ``I don't know'' option, selecting a random answer, choosing always the same option in a response scale, abandoning the survey without completing it, etc.

We measured the satisficing effect through the differentiation response index from \cite{mccarty2000measurement}, which quantifies the level of heterogeneity between the answers given by the same user. In our case, all questions had a closed set of answer options that never included a no-opinion alternative and all respondents completed the survey to complete the crowdsourcing task, so the differentiation index was the most suitable metrics to estimate the response quality. 

In a comparison between face-to-face and web surveying, \cite{heerwegh2008face} found that Web respondents show a lower differentiation response index with respect to face-to-face interviewing; similarly, we want to test if our conversational approach, being inspired by inter-personal interaction, induces a higher differentiation index than a traditional Web surveying method.

Table~\ref{tab:satisficing} shows the experimental results of the differentiation index in the three versions of the survey (TF, CF and CI). We also performed a pairwise ANOVA to test for the difference between the mean differentiation indexes, and  all three differences resulted to be statistically significant.

\begin{table}[htb]
\begin{center}
\begin{tabular}{lccc}
\toprule
Differentiation     & Traditional & Conversational & Conversational \\
index               & Formal      & Formal         & Informal       \\
\midrule                                                             
mean                & 0.65        & 0.62           & 0.68           \\
standard deviation  & 0.09        & 0.09           & 0.07           \\
\bottomrule      
\end{tabular}
\end{center}
\caption{Differentiation index between answers: a higher value means a lower level of satisficing (N=100 for each survey).}
\label{tab:satisficing}
\end{table}

The above results therefore show that the conversational approach with informal formulation (CI survey) indeed leads to a higher differentiation index, which can be interpreted as a lower degree of satisficing as suggested by \cite{krosnick1991response}. It may appear surprising though that the differentiation index of the CF survey is significantly lower that the respective measure of the TF survey, but this is also in line with the findings by \cite{kim2019comparing}. 

We interpret this outcome in relation to the interaction experience: in the conversational tool, the survey interface scrolls automatically so to let the answer options always at the bottom of the page; this allows the respondent to continue clicking in the same area without manually scrolling or moving the mouse, as they do in the respective SurveyMonkey version. Our conclusion is therefore that the conversational approach can lead to a higher response quality possibly reducing the satisficing effect, provided that it is accompanied by a rich and interactive experience that solicits the answer differentiation by increasing the ``burden'' of answering. 

\subsection{Conclusions on method effectiveness}\label{sub:eff}
Our experimental results show that our conversational survey approach displays a very similar level of reliability with respect to its traditional counterpart, both in terms of item consistency and with respect to response equivalence in terms of inter-rater reliability. We can therefore say that our approach is at least as reliable as a more classic survey method. Moreover, the response quality results prove that a conversational form can lead to an improvement with respect to traditional approaches by helping to reduce the satisficing effect, when accompanied by an informal and more human-like interaction style.

\section{Conclusions}\label{sec:concl}
In this paper, we presented our conversational survey approach and the results of our evaluation. From the compiling user point of view, our method shows a clear advantage in terms of user experience and acceptance, because the large majority of our experimenters expressed a strong preference for the conversational interface. From the questionnaire data collection point of view, we also demonstrated that our approach performs well in terms of both reliability and response quality, aspiring to become a credible alternative or even a preferred method with respect to traditional tools.

We are aware that this represents only the first step towards a full evaluation of the conversational survey approach. We believe, however, that the presented results have value, especially when attention is given to the compiling activity: an improved user experience can lead to more engaged participants to data gathering and, as a consequence, to more reliable collected information with higher response quality. 

We argue that the main implication for survey research is that questionnaire design methodologies should include an additional step to identify the best approach to engage respondents and ensure an enjoyable survey administration. The formulation of survey items should be oriented not only to be clear and unambiguous, but also to be interesting for respondents: \cite{saris2010comparing} showed that substituting an agree-disagree scale with item-specific response options leads to superior data quality; our conversational approach brings this change in formulation further, by suggesting the use of more colloquial and informal articulation, more similar to a qualitative research method even when collecting quantitative data. Moreover, even if people are quite used to traditional means to fill in a survey, the choice of the questionnaire administration channel should be oriented to maximize the collected data quality; our proposed toolkit shows that a chat channel can be a valuable substitute of a more classic interface, which can lead to notable and positive effects (increased compilation time and more focused attention). 

Our next steps are directed not only to gather more evidence of user experience and method effectiveness. We plan also to test the conversation editor tool with questionnaire designers to assess the impact of this innovative way of creating surveys on the investigation process. Informal discussion with psychologists, for example, raised the research question whether our tool is a means to a completely new family of survey methods, rather than a substitute of the existing ones. We are indeed aware that, to create a consistent and enjoyable conversation, additional effort is required at design time and that storytelling capabilities are not usually considered mandatory among survey designer skills. Still, we believe that the  results we offered in this paper show that adopting our conversational survey approach is worth the effort.

\section*{Acknowledgment}
\small The research presented in this paper was partially supported by the Bankable project (id 18165), co-funded by EIT Digital under the Digital Finance Action Line, and the ACTION project (grant agreement number 824603), co-funded by the European Commission under the Horizon 2020 Framework Programme. We would like to thank the 700 participants recruited through the Prolific crowdsourcing platform for their valuable contribution to our experimentation. Special thanks go also to the colleagues that helped up in the CONEY toolkit design and development as well as in the evaluation support: Lorenzo Bernaschina, Fabiano Rivolta, Damiano Scandolari, Mario Scrocca, Emanuela Carlino and Davide Stenner.

  \bibliographystyle{elsarticle-harv} 
  \bibliography{biblio}

\appendix

\section{Mobile banking questionnaire: traditional survey content}\label{app:traditional}
The mobile banking traditional survey, taken from \cite{kim2009understanding}, consists of 21 items representing 7 different latent variables, as per the following list. For each item, the user was given a 1-5 Likert scale of agreement (Strongly disagree, Disagree, Neutral, Agree, Strongly agree) to express his/her own opinion on the subject. All items were optional and respondents were instructed to leave an item blank if they did not have an opinion or if they found the question unclear. However, only a handful of items were actually left blank, with no impact on the overall data analysis presented in this paper. 
 
\begin{itemizesmall}
	\item \{Relative benefits\} Mobile banking has more advantages than Internet or off-line banking because services are not limited by location.
	\item \{Relative benefits\} Mobile banking is more convenient than Internet or off-line banking.
	\item \{Relative benefits\} Mobile banking is more efficient than Internet or off-line banking.
	\item \{Relative benefits\} Mobile banking is more effective than Internet or off-line banking in managing a bank account.
	\item \{Propensity to trust\} I am cautious in using new technologies to do my work.
	\item \{Propensity to trust\} If possible, it is better to avoid using new technologies for financial transactions.
	\item \{Propensity to trust\} When using a new technology, I have to be careful until I see the evidence of a technology provider's trustworthiness.
	\item \{Perceived structural assurance\} Mobile banking firms guarantee compensation for monetary losses that might occur during service usage.
	\item \{Perceived structural assurance\} Mobile banking firms guarantee the protection of customers' personal information.
	\item \{Perceived structural assurance\} Mobile banking firms publish a policy on the protection of transactional data.
	\item \{Perceived structural assurance\} Mobile banking firms publish a policy on customer protection from accidents.
	\item \{Firm reputation (mobile)\} My mobile telephone operator has a good reputation.
	\item \{Firm reputation (mobile)\} My mobile telephone operator is recognized widely.
	\item \{Firm reputation (mobile)\} My mobile telephone operator offers good services.
	\item \{Firm reputation (bank)\} My bank has a good reputation.
	\item \{Firm reputation (bank)\} My bank is recognized widely.
	\item \{Firm reputation (bank)\} My bank offers good services.
	\item \{Initial trust\} Mobile banking always provides accurate financial services.
	\item \{Initial trust\} Mobile banking always provides reliable financial services.
	\item \{Initial trust\} Mobile banking always provides safe financial services.
	\item \{Usage intention\} I intend to use mobile banking.
\end{itemizesmall}

\section{Mobile banking questionnaire: conversational survey content}\label{app:conversation}
The conversational survey is a pre-defined dialogue in which the system engages the compiler in a chat asking him/her the questionnaire questions. The dialogue is made up of a list of building blocks, questions, answers and textual content (colloquial parts of the conversation); those blocks are connected in a sequence that, however, can have branches in relation to the choices the user makes during the chat. 

\begin{figure}[b]
	\centering
		\includegraphics[width=1.00\textwidth]{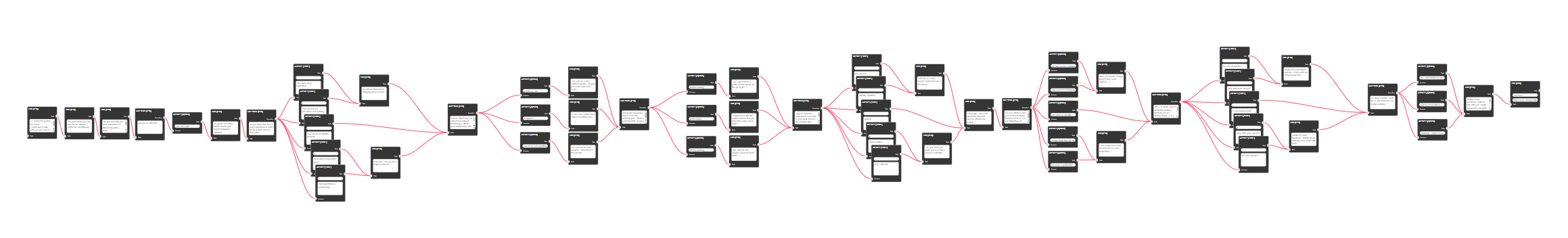}
	\caption{Entire conversational survey flows in terms of sequence of text/question/ answer blocks with branches.}
	\label{fig:conv-all}
\end{figure}

\begin{figure}[t]
	\centering
		\includegraphics[width=0.90\textwidth]{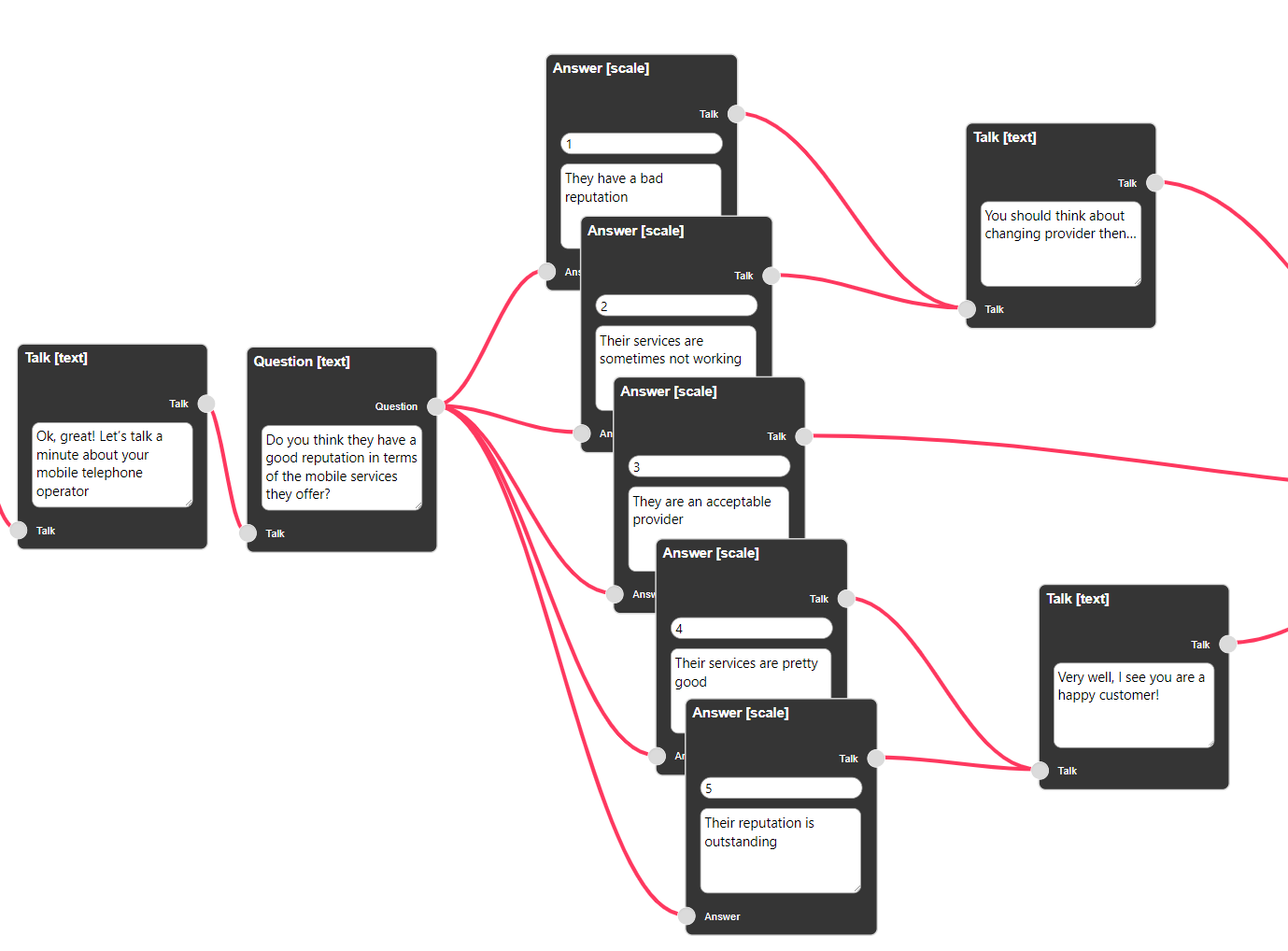}
	\caption{Zoom-in on the mobile firm reputation question of the conversational survey.}
	\label{fig:conv-zoom-in}
\end{figure}

In the following we report the content of the mobile banking conversational survey and the way this ``scripted'' dialogue appears as flow of building blocks in the editor (Figures \ref{fig:conv-all} and \ref{fig:conv-zoom-in}). For the validation of the items, we performed a pre-test of the questionnaire with two participants.

\begin{itemizesmall}
	\item \{text\} Hi! Thanks for joining me today! I would like to talk with you and ask you a few questions about mobile banking
	\item \{text\} I'll assume that you own and use an Internet-connected smartphone
	\item \{text\} Oh, and also that you have experience of interacting with a bank (think of the bank you interact with more frequently, if you have multiple accounts)
	\item \{question\} Are you ok with this?
	\item \{answer\} Sure, let's start!
	\item \{text\} Ok, great! Let's talk a minute about your mobile telephone operator
	\item \{question: Firm reputation (mobile)\} Do you think they have a good reputation in terms of the mobile services they offer?
	\item \{answer, value: 1\} They have a bad reputation 
	\item \{answer, value: 2\} Their services are sometimes not working 
	\item \{answer, value: 3\} They are an acceptable provider 
	\item \{answer, value: 4\} Their services are pretty good 
	\item \{answer, value: 5\} Their reputation is outstanding 
	\item \{text, if answer 1 or 2\} You should think about changing provider then...
	\item \{text, if answer 4 or 5\} Very well, I see you are a happy customer!
	\item \{question: Propensity to trust\}  It seems that today we cannot live without technology!  What's your relationship with it? Do you easily adopt new technologies or apps as soon as they are out?
	\item \{answer, value: 1\} As little as I can... 
	\item \{answer, value: 3\} It depends... 
	\item \{answer, value: 5\} I like to try out everything! 
	\item \{text, if answer 1\} I see you are a very cautious person... maybe you could dare a bit more :-)
	\item \{text, if answer 3\} It's ok to be careful when trying something new!
	\item \{text, if answer 5\} I see you are an early adopter... innovation is the answer!
	\item \{question: Relative benefits\}  Also banks nowadays rely a lot on new technologies... What is your favourite channel to interact with your bank?
	\item \{answer, value: 5\} Mobile banking 
	\item \{answer, value: 3\} Web banking 
	\item \{answer, value: 1\} Off-line banking 
	\item \{text, if answer 5\} I see, you must be a globe-trotter always on the go, right? ;-)
	\item \{text, if answer 3\} I suppose you like the digital channel, but you still prefer a big screen, right? ;-)
	\item \{text, if answer 1\} Well, that for sure ensures a human touch, right? ;-)
	\item \{text\} Lately it's become mainstream accessing to your bank services via a mobile app.
	\item \{question: Perceived structural assurance\} Tell me how confident you are that your bank will protect your data and your transactions during mobile banking activities
	\item \{answer, value: 1\} Very unsure 
	\item \{answer, value: 2\} Partially doubtful 
	\item \{answer, value: 3\} Neutral
	\item \{answer, value: 4\} Quite positive 
	\item \{answer, value: 5\} Very confident 
	\item \{text, if answer 1 or 2\} Don't be so scared, secure connections are a fact today!
	\item \{text, if answer 4 or 5\} I see you trust your bank, you must have chosen it carefully!
	\item \{text\} Personally, I care a lot about the financial services offered via mobile
	\item \{question: Initial trust\} Are you satisfied with your mobile banking financial services? Do you think they are accurate, reliable and safe enough?
	\item \{answer, value: 1\} No, they are unacceptable
	\item \{answer, value: 2\} They need lots of improvements
	\item \{answer, value: 3\} Somewhat acceptable
	\item \{answer, value: 4\} I mostly think they are ok
	\item \{answer, value: 5\} Yes, close to perfection!
	\item \{text, if answer 1 or 2\} Wow, you should change bank if that's your opinion... :-(
	\item \{text, if answer 4 or 5\} Cool, maybe you could recommend me your bank then! :-)
	\item \{question: Firm reputation (bank)\} Now, I'm really curious about the banking institute you are talking about... Is it a well-established bank? Are customers usually satisfied by their services?
	\item \{answer, value: 1\} No, very unsatisfied 
	\item \{answer, value: 2\} They offer poor services
	\item \{answer, value: 3\} It's an average bank 
	\item \{answer, value: 4\} They offer good services 
	\item \{answer, value: 5\} Yes, very satisfied!
	\item \{text, if answer 1 or 2\} Thanks for your honest opinion, I'll stick with my current bank then...
	\item \{text, if answer 4 or 5\} Thanks for your feedback, I'll think about opening an account with them!
	\item \{question: Usage intention\}  OK, final question: All in all, do you intend to use mobile banking?
	\item \{answer, value: 1\} No, I don't intend to use it
	\item \{answer, value: 3\} I will use it sometimes
	\item \{answer, value: 5\} Definitely, I intend to use it
	\item \{text\} Thanks, it was extremely useful to talk with you! I really appreciate you spent some time chatting with me.  Before you go, would you mind evaluating our conversation? Click here
\end{itemizesmall}

\section{Motivation to participate questionnaire: traditional/ conversational formal survey content}\label{app:formal}
We designed the survey to investigate the motivation to participate to crowdsourcing by taking inspiration from the similar survey on citizen science by \cite{levontin2018motivation}, which in turn is based on the theory of human values by \cite{schwartz2005basic}. First we selected a subset of 10 latent variables (self-direction, stimulation, routine, hedonism, achievement, power, belongingness, conformity, benevolence and universalism) and chose two survey items for each variable; all those items were expressed as statements for which the user had to choose his/her level of agreement on a 5-point Likert scale. Additionally we included one item to let the respondents express their global level of motivation, again on a 5-point Likert scale. For the validation of the items, we performed a pre-test of the questionnaire with two participants for each version. The verbalization below is the formal-style formulation adopted in both the respective traditional and conversational experiments (referred to as TF and CF surveys in the paper).

\begin{itemizesmall}
	\item \{Self-direction\} I want to learn
	\item \{Self-direction\} I am interested in crowdsourcing
	\item \{Stimulation\} I want to do something new
	\item \{Stimulation\} I strive to challenge myself
	\item \{Routine\} I was doing crowdsourcing anyway
	\item \{Routine\} I am a regular participant in crowdsourcing campaigns
	\item \{Hedonism\} Participating makes me feel good about myself
	\item \{Hedonism\} I am passionate about crowdsourcing
	\item \{Achievement\} It is an opportunity to perform better than others
	\item \{Achievement\} I want to do something meaningful
	\item \{Power\} I want to gain recognition and status
	\item \{Power\} I expect something in return
	\item \{Belongingness\} I want to meet people with similar interests
	\item \{Belongingness\} I want to feel part of something worthwhile
	\item \{Conformity\} Other people I know are participating
	\item \{Conformity\} I was requested to participate by somebody
	\item \{Benevolence\} It is a good thing to do
	\item \{Benevolence\} I want to contribute to scientific research
	\item \{Universalism\} I want to make scientific knowledge more accessible
	\item \{Universalism\} I want to raise public awareness to the topic of crowdsourcing campaigns
	\item \{Global motivation\} How much are you motivated in participating to crowdsourcing campaigns?
\end{itemizesmall}

\section{Motivation to participate questionnaire: conversational informal survey content}\label{app:informal}
We designed the informal version of the questionnaire presented in Appendix~\ref{app:formal} specifically for administration through the conversational tool. Each item was re-formulated, both adopting a more casual and colloquial style and differentiating the answer options between questions, so to diversify the compilation experience. Therefore, this version also includes two items for each of the ``values'' latent variables and one final item for the global motivation. Also in this case, we performed a pre-test of the questionnaire with two participants for the validation of the items. The verbalization below is the informal-style formulation adopted in the respective conversational experiment (referred to as CI survey in the paper). To understand the survey element, please refer to the general explanation offered in Appendix~\ref{app:conversation}.

\begin{itemizesmall}
	\item \{text\} Hello! I would like to ask you some questions about your motivations in participating to crowdsourcing campaigns. Think about your experience with Prolific while answering the questions.
	
	\item \{text\} Great, we can start with the questionnaire!
	
	\item \{question: self-direction\} How much do you expect to learn from your participation crowdsourcing campaigns?
	\item \{answer, type: star-rating, value: 5\} *****
	\item \{answer, type: star-rating, value: 4\} ****
	\item \{answer, type: star-rating, value: 3\} ***
	\item \{answer, type: star-rating, value: 2\} **
	\item \{answer, type: star-rating, value: 1\} *
	
	\item \{question: self-direction\} Are you interested in crowdsourcing?
	\item \{answer, type: emoji, value: 5\} Very curious
	\item \{answer, type: emoji, value: 4\} Curious
	\item \{answer, type: emoji, value: 3\} Neutral
	\item \{answer, type: emoji, value: 2\} Care very little
	\item \{answer, type: emoji, value: 1\} Don't care
	
	\item \{question: stimulation\} Did you join crowdsourcing campaigns to have the possibility to do something new?
	\item \{answer, type: star-rating, value: 5\} *****
	\item \{answer, type: star-rating, value: 4\} ****
	\item \{answer, type: star-rating, value: 3\} ***
	\item \{answer, type: star-rating, value: 2\} **
	\item \{answer, type: star-rating, value: 1\} *
	
	\item \{question: stimulation\}  Do you think your participation is an opportunity to challenge yourself?
	\item \{answer, type: slide, value: 5\} Exactly
	\item \{answer, type: slide, value: 4\} Partially
	\item \{answer, type: slide, value: 3\} Not influenced
	\item \{answer, type: slide, value: 2\} A bit
	\item \{answer, type: slide, value: 1\} Not at all
	
	\item \{question: routine\} Have you ever done crowdsourcing campaigns before?
	\item \{answer, type: options, value: 5\} Weekly or more often
	\item \{answer, type: options, value: 4\} Monthly
	\item \{answer, type: options, value: 3\} Yearly
	\item \{answer, type: options, value: 2\} Once/Twice
	\item \{answer, type: options, value: 1\} Never
	
	\item \{question: routine\}  How regularly do you participate to crowdsourcing campaigns?
	\item \{answer, type: star-rating, value: 5\} *****
	\item \{answer, type: star-rating, value: 4\} ****
	\item \{answer, type: star-rating, value: 3\} ***
	\item \{answer, type: star-rating, value: 2\} **
	\item \{answer, type: star-rating, value: 1\} *
	
	\item \{text\} You are doing great! Let's proceed with some different questions
	
	\item \{question:hedonism\} Does your participation to crowdsourcing campaigns make you feel good about yourself?
	\item \{answer, type: star-rating, value: 5\} *****
	\item \{answer, type: star-rating, value: 4\} ****
	\item \{answer, type: star-rating, value: 3\} ***
	\item \{answer, type: star-rating, value: 2\} **
	\item \{answer, type: star-rating, value: 1\} *

	\item \{question:hedonism\} How passionate are you about the crowdsourcing initiative?
	\item \{answer, type: star-rating, value: 5\} *****
	\item \{answer, type: star-rating, value: 4\} ****
	\item \{answer, type: star-rating, value: 3\} ***
	\item \{answer, type: star-rating, value: 2\} **
	\item \{answer, type: star-rating, value: 1\} *
	
	\item \{question: achievement\}  Does the participation to crowdsourcing campaigns represent an opportunity for you to perform better than others in some respects?
	\item \{answer, type: slide, value: 5\} Yes
	\item \{answer, type: slide, value: 4\} Yes a little
	\item \{answer, type: slide, value: 3\} Neutral
	\item \{answer, type: slide, value: 2\} Not a lot
	\item \{answer, type: slide, value: 1\} Not at all
	
	\item \{question: achievement\}  Does your participation to crowdsourcing campaigns represent an opportunity to do something meaningful?
	\item \{answer, type: slide, value: 5\} Yes
	\item \{answer, type: slide, value: 4\} Yes a little
	\item \{answer, type: slide, value: 3\} Neutral
	\item \{answer, type: slide, value: 2\} Not a lot
	\item \{answer, type: slide, value: 1\} Not at all
	
	\item \{question: power\} Do you believe your participation allows you to gain recognition and status?
	\item \{answer, type: emoji, value: 5\} Yes
	\item \{answer, type: emoji, value: 4\} Yes a little
	\item \{answer, type: emoji, value: 3\} Neutral
	\item \{answer, type: emoji, value: 2\} Not a lot
	\item \{answer, type: emoji, value: 1\} Not at all
	
	\item \{question: power\} Do you expect something in return from your participation to crowdsourcing campaigns?
	\item \{answer, type: options, value: 5\} Great payoff
	\item \{answer, type: options, value: 4\} Considerable payoff
	\item \{answer, type: options, value: 3\} Something
	\item \{answer, type: options, value: 2\} Almost nothing
	\item \{answer, type: options, value: 1\} Nothing
	
	\item \{text\} Awesome! We've almost done, so don't leave me now!
	
	\item \{question: belongingness\} Is your participation to crowdsourcing campaigns influenced by the desire to meet people with similar interests?
	\item \{answer, type: options, value: 5\} Strongly influenced
	\item \{answer, type: options, value: 4\} Influenced
	\item \{answer, type: options, value: 3\} Neutral
	\item \{answer, type: options, value: 2\} A little
	\item \{answer, type: options, value: 1\} Not at all
	
	\item \{question: belongingness\} By joining crowdsourcing campaigns, do you feel part of something worthwhile?
	\item \{answer, type: emoji, value: 5\} Yes
	\item \{answer, type: emoji, value: 4\} Yes a little
	\item \{answer, type: emoji, value: 3\} Neutral
	\item \{answer, type: emoji, value: 2\} Not a lot
	\item \{answer, type: emoji, value: 1\} Not at all
	
	\item \{question: conformity\} Do you know other people participating to crowdsourcing campaigns?
	\item \{answer, type: options, value: 5\} A number of people
	\item \{answer, type: options, value: 4\} A quite big number of people
	\item \{answer, type: options, value: 3\} A few participants
	\item \{answer, type: options, value: 2\} Only one participant
	\item \{answer, type: options, value: 1\} No one
	
	\item \{question: conformity\} To what degree were you obliged to participate?
	\item \{answer, type: options, value: 5\} It was strongly mandatory for me
	\item \{answer, type: options, value: 4\} It was mandatory for me
	\item \{answer, type: options, value: 3\} I was suggested to
	\item \{answer, type: options, value: 2\} It was partially my own choice
	\item \{answer, type: options, value: 1\} It was my own choice
	
	\item \{text\} Well done! I have only the last set of questions for you
	
	\item \{question: benevolence\} How much do you see your participation in the crowdsourcing campaigns as a good thing to do?
	\item \{answer, type: star-rating, value: 5\} *****
	\item \{answer, type: star-rating, value: 4\} ****
	\item \{answer, type: star-rating, value: 3\} ***
	\item \{answer, type: star-rating, value: 2\} **
	\item \{answer, type: star-rating, value: 1\} *
	
	\item \{question: benevolence\} Do you participate to contribute and help the scientific research?
	\item \{answer, type: star-rating, value: 5\} *****
	\item \{answer, type: star-rating, value: 4\} ****
	\item \{answer, type: star-rating, value: 3\} ***
	\item \{answer, type: star-rating, value: 2\} **
	\item \{answer, type: star-rating, value: 1\} *
	
	\item \{question: universalism\} Do you participate for the possibility to make data about crowdsourcing campaigns more accessible?
	\item \{answer, type: options, value: 5\} Definitely
	\item \{answer, type: options, value: 4\} Mostly
	\item \{answer, type: options, value: 3\} Partially
	\item \{answer, type: options, value: 2\} Mostly for other reasons
	\item \{answer, type: options, value: 1\} Not at all
	
	\item \{question: universalism\} How much do you see your participation as a possibility to raise public awareness to the topic of the crowdsourcing campaign?
	\item \{answer, type: star-rating, value: 5\} *****
	\item \{answer, type: star-rating, value: 4\} ****
	\item \{answer, type: star-rating, value: 3\} ***
	\item \{answer, type: star-rating, value: 2\} **
	\item \{answer, type: star-rating, value: 1\} *
	
	\item \{question: global motivation\} How much are you motivated in participating to crowdsourcing campaigns?
	\item \{answer, type: star-rating, value: 5\} *****
	\item \{answer, type: star-rating, value: 4\} ****
	\item \{answer, type: star-rating, value: 3\} ***
	\item \{answer, type: star-rating, value: 2\} **
	\item \{answer, type: star-rating, value: 1\} *
	
	\item \{question\} In your own words, what is the main motivation why you decided to participate to crowdsourcing campaigns?
	
	\item \{text\} Thank you for your answers and for your time!
	
	\item \{text\} Bye! Keep contributing to crowdsourcing!

\end{itemizesmall}

\end{document}